\documentclass[12pt,a4paper]{article}
\usepackage{amssymb}
\usepackage{graphicx}
\usepackage{amsmath}

\begin{document}

\title{Effects of the resonant magnetic perturbation on turbulent transport}
\author{{\small {M. Vlad, F.Spineanu}} \\
\textit{{\small {National Institute of Laser, Plasma and Radiation Physics,
Atomi\c{s}tilor}}} \\
\textit{{\small {409, 077125 Magurele, Bucharest, Romania}}}}
\date{}
\maketitle

\begin{abstract}
The effects of the resonant magnetic perturbations (RMPs) on the turbulent
transport are analyzed in the framework of the test particle approach using
a semi-analytical method. The normalized RMP amplitude $P_{b}$ extends on a
large range, from the present experiments to ITER conditions. The results
are in agreement with the experiments at small $P_{b}.$ The predictions for
ITER strongly depend on the type of turbulent transport. A very strong
increase of the turbulent transport is obtained in the nonlinear regime,
while the effects of the RMPs are much weaker for the quasilinear transport.

\end{abstract}

\section{Introduction}

The interaction of resonant magnetic perturbations (RMP) with tokamak
plasmas determines complex physical precesses that are intensively studied
in view of ITER, as seen in the very recent review paper \cite{review}. The
goal of suppression or mitigation of the edge localized mode \cite{elm1}, 
\cite{elm2} has been achieved in present devices, but there are many aspects
that are not understood.

In particular, the studies of the effects of the RMPs on turbulence and
transport are in progress. The experiments have clearly shown the increase
of turbulence and of the transport in the presence of RMPs in tokamak
plasmas \cite{turb4}-\cite{turb-L} and in other configurations as RFX \cite%
{RFX} and LHD \cite{LHD}. Theoretical models and numerical simulations have
confirmed these effects (\cite{Cad2014}-\cite{turb3} and the references
there in)

The aim of this paper is to evaluate the direct effects on the RMPs on the
turbulent transport as function of the parameters of the turbulence. This is
a complementary approach to the selfconsistent numerical simulations, which
determine the characteristics of the turbulence and the transport as
function of the macroscopic plasma parameters (the gradients, heating,
etc.). We present a test particle study, which is expected to bring a
different perspective that could contribute to the understanding of the
complex interaction process.

The transport regimes for the ions and for the electrons (including
multi-scale effects) were recently studied \cite{VS2013}, \cite{VS2015}. We
determine here, using the same semi-analytical method \cite{V1998}, the
effects of the RMPs in each transport regime.

The paper is organized as follows. The model is described in section $2.$ It
is a multi-stochastic process determined by the turbulence, the RMPs and
particle collisions. The derivation of the statistical solution for the time
dependent transport coefficients is presented in Section $3$. We show first
that the multi-stochastic process that describe collisional particle
transport can be reduced to the transport in stochastic effective velocity
field and we determine its Eulerian correlation. The effects of the RMPs on
the diffusion coefficients and on the pinch velocity determined by the
gradient of the toroidal magnetic field are presented and analyzed in
Section $4.$ Section $5$\ contains the summary and the conclusions drawn
from this work.

\section{The model}

The test particle approach of transport is based on the evaluation of the
Lagrangian velocity correlation (LVC) as a function of the characteristics
of the turbulence represented by the Eulerian correlation (EC) or spectrum
of the fluctuating potential. The LVC\ is a time-dependent function $L(t)$
that defines the Lagrangian decorrelation time $\tau _{L}.$\ It is the
characteristic decay time of the LVC and it measures the statistical memory
of the stochastic process. The integral of $L(t)$\ is the time dependent
diffusion coefficient \cite{Taylor}. The LVC is defined by the statistical
average of the trajectories calculated for short times $t\lesssim \tau _{L}.$
The test particle approach can be applied when the typical displacements
produced during $\tau ,$ which are of the order of the correlation length of
the turbulence, are much smaller than the characteristic lengths of the
temperature and density \cite{B}.

We consider a homogeneous and stationary turbulent plasma represented by a
stochastic potential $\phi (\mathbf{x},z,t),$ where $\mathbf{x}=(x,y)$ are
the coordinates in the plane perpendicular to the confining magnetic field $B%
\mathbf{e}_{z}$ (with $x$\ in the radial and $y$\ in the poloidal
directions) and $z$ is the parallel coordinate. The trajectories of the
guiding centers are solutions of 
\begin{eqnarray}
\frac{d\mathbf{x}}{dt} &=&-\frac{\mathbf{\nabla }\phi \left( x,y^{\prime
},z,t\right) \times \mathbf{e}_{z}}{B}+\dfrac{\mathbf{b}(\mathbf{x},z)}{B}%
\eta _{\parallel }(t),  \label{1-x} \\
\frac{dz}{dt} &=&\eta _{\parallel }(t),  \label{1-z}
\end{eqnarray}%
where the first term is the $\mathbf{E}\times \mathbf{B}$ drift, the second
term is the velocity determined by the motion along the perturbed magnetic
field, $\mathbf{b}(\mathbf{x},z)$ is the perpendicular magnetic field
produced by RMP coils, and $\eta _{\parallel }(t)$ is the parallel
collisional velocity. The perpendicular collisional velocity is negligible
in Eq. (\ref{1-x}).

The EC of $\phi ,$ which is an input function in test particle studies, is
modeled using the results obtained in the numerical simulations for the ion
gradient driven turbulence \cite{nITG1}-\cite{nITG2} or for the trapped
electron modes \cite{TEM}. The modes with zero poloidal wave number $k_{y}=0$
are stable for both types of turbulence, which leads to a special shape of
the EC that has the poloidal integral equal to zero. The model presented in 
\cite{VS2013} is used 
\begin{eqnarray}
E(\mathbf{x},z,t) &\equiv &\left\langle \phi (\mathbf{0},0,0)~\phi (\mathbf{x%
},z,t)\right\rangle _{\phi }  \label{rEC} \\
&=&\Phi ^{2}\exp \left( -\frac{x^{2}}{2\lambda _{x}^{2}}-\frac{z^{2}}{%
2\lambda _{z}^{2}}-\frac{t}{\tau _{c}}\right) \partial _{y^{\prime }}\left[
\exp \left( -\frac{y^{\prime \ 2}}{2\lambda _{y}^{2}}\right) y^{\prime }%
\right] ,  \notag
\end{eqnarray}%
where $\Phi $ is the amplitude of the potential fluctuations, $\lambda _{i},$
$i=x,y,z$\ are the correlation lengths along the three directions and $\tau
_{c}$\ is the correlation time. The derivative $\partial _{y^{\prime
}}\equiv \partial /\partial y^{\prime }$ ensure that the poloidal integral
is zero. The poloidal drift of the potential with the effective diamagnetic
velocity $V_{d}$ is represented by $y^{\prime }=y-V_{d}t$ in Eq. (\ref{rEC}%
). The correlations of the components of the $\mathbf{E}\times \mathbf{B}$
stochastic velocity $v_{i}=-\varepsilon _{ij}\partial _{j}\phi (\mathbf{x}%
,z,t)$ are obtained from the EC of the potential%
\begin{eqnarray}
C_{ij}(\mathbf{x},z,t) &\equiv &\left\langle v_{i}(\mathbf{0},0,0)~v_{j}(%
\mathbf{x},z,t)\right\rangle _{\phi }  \notag \\
&=&-\varepsilon _{ik}\varepsilon _{jl}\partial _{k}\partial _{l}E(\mathbf{x}%
,z,t),  \label{vEC}
\end{eqnarray}%
where $\varepsilon _{12}=1,$ $\varepsilon _{21}=-1,$ $\varepsilon _{ii}=0,$ $%
\partial _{1}=\partial /\partial x,$\ and $\partial _{2}=\partial /\partial
y.$ \ \ 

The confining magnetic field is 
\begin{equation}
B(x)=B_{0}\exp \left( -\dfrac{x}{R}\right) \simeq B_{0}\left( 1-\dfrac{x}{R}%
\right) ,  \label{Bt}
\end{equation}%
where $R$ is the major radius of the plasma. The space variation of the
confining magnetic field is included in the model through a small gradient
in the radial direction, $\mathbf{\nabla }B\simeq B_{0}/R\mathbf{e}_{x}.$

The magnetic perturbations can be resonant (with one dominant mode) or
stochastic (with a spectrum of modes of comparable amplitudes) \cite{spectr}%
. Considering the averages related to the analysis of the turbulent
transport that essentially are space averages over the initial conditions,
the magnetic field $\mathbf{b}(\mathbf{x},z)$ contributes to particle
trajectories as a stochastic function in both cases.\ The magnetic field $%
\mathbf{b}(\mathbf{x},z)$ is represented by a stochastic function that has
macroscopic correlation lengths. They are determined by the number and the
configuration of the RMP coils and are much larger than the perpendicular
correlation length of the turbulence $\lambda _{x},~\lambda _{y}$, which are
of the order of the ion Larmor radius $\rho _{i}$. Since the trajectories
have to be statistically analyzed only for short times that corresponds to
displacements at turbulence space scale, the radial and the poloidal
variation of $\mathbf{b}(\mathbf{x},z)$ can be neglected. However we take
into account the $y$\ dependence of $\mathbf{b}$ due to the poloidal
component of the unperturbed magnetic field, which determines the rotation
of the magnetic lines on the magnetic surfaces and leads to poloidal
decorrelation. The toroidal correlation length of $\mathbf{b}$ is smaller
than $\lambda _{z},$ which is of the order of $2\pi R$. We also neglect the
poloidal component of $\mathbf{b,}$\ which is small and is expected to have
a weak effect on the radial diffusion coefficient. Thus, the magnetic
perturbation is approximated with $\mathbf{b}=b(y,z)\mathbf{e}_{x}.$\ The EC
of the magnetic field is modeled by%
\begin{equation}
C(y,z)\equiv \left\langle b(0,0)~b(y,z)\right\rangle _{b}=\beta ^{2}\exp
\left( -\frac{y^{2}}{2\Lambda _{y}^{2}}--\frac{z^{2}}{2\Lambda _{z}^{2}}%
\right) ,  \label{ECmf}
\end{equation}%
where $\beta $\ is the amplitude of the RMPs and $\Lambda _{y},$ $\Lambda
_{z}$\ are the correlation lengths. They are essentially determined by the
number of coils $n$\ and $m$\ in the toroidal and poloidal directions,
respectively%
\begin{equation}
\Lambda _{z}=\frac{2\pi R}{n},\ \Lambda _{y}=\frac{\alpha ~a}{m},
\label{lambda}
\end{equation}%
\ \ where $a$ is the minor radius and $\alpha $\ is the angle of poloidal
extension of the coils.\ \ 

The EC of the collisional velocity $\eta _{\parallel }(t)$ is 
\begin{equation}
C_{\parallel }(t)=\left\langle \eta _{\parallel }(0)\eta _{\parallel
}(t)\right\rangle _{\parallel }=\chi _{\parallel }\nu ~\exp (-\nu \left\vert
t\right\vert ),  \label{cpar}
\end{equation}%
where $\nu $ is the frequency of collisions, $\chi _{\parallel }=\lambda
_{mfp}^{2}\nu $ is the parallel diffusivity, $\lambda _{mfp}$ is the mean
free path, and $\left\langle {}\right\rangle _{\parallel }$\ is the
statistical average.

Dimensionless variables are introduced. The perpendicular displacements, the
correlation lengths ($\lambda _{x},$ $\lambda _{y},$ $\Lambda _{y})$\ and
the gradient length $R$ are normalized with the ion Larmor radius $\rho _{i}$%
. The parallel displacements $z,$ $\lambda _{z},$ $\Lambda _{z}$ are divided
by $L_{T_{i}},$ the gradient length of the ion temperature $T_{i}$. The unit
of time and of $\tau _{c}$ is $\tau _{0}=L_{T_{i}}/v_{T_{i}},$ where $%
v_{T_{i}}=\sqrt{T_{i}/m_{i}}$\ is the thermal velocity of the ions and $%
m_{i} $\ is their mass.\ The collisional velocity is normalized with the
amplitude $\sqrt{\chi _{\parallel }\nu }=v_{T_{i}}.$ The units for potential
and for the magnetic field are $\Phi $ and $\beta ,$ respectively. Using the
same symbols for the dimensionless variables, the equations of motion in the
reference system that moves with the potential are%
\begin{eqnarray}
\frac{dx}{dt} &=&-P_{\phi }\exp (\frac{x}{\overline{R}})\mathbf{~\partial }%
_{y}\phi \left( x,y,z,t\right) +P_{b}\exp (\frac{x}{\overline{R}})b(y,z)\eta
_{\parallel }(t)  \label{1p-x} \\
\frac{dy}{dt} &=&P_{\phi }\exp (\frac{x}{\overline{R}})\mathbf{~\partial }%
_{x}\phi \left( x,y,z,t\right) +\overline{V}_{d}  \label{1p-y} \\
\frac{dz}{dt} &=&\eta _{\parallel }(t)  \label{1p-z}
\end{eqnarray}%
where $\overline{V}_{d}=V_{d}/V_{\ast }$\ with the diamagnetic velocity $%
V_{\ast }=\rho _{i}~v_{T_{i}}/L_{T_{i}}.$ All the stochastic fields in Eqs. (%
\ref{1p-x}-\ref{1p-z}) have the amplitudes equal to one. The correlation
function of the normalized collisional velocity is 
\begin{equation}
C_{\parallel }(t)=\exp (-P_{c}\left\vert t\right\vert ).  \label{Rt}
\end{equation}%
Three dimensionless parameters appear in the normalized equations (\ref{1p-x}%
-\ref{1p-z})%
\begin{equation}
P_{\phi }\equiv \frac{\Phi }{B_{0}\rho _{i}V_{\ast }}=\frac{e\Phi }{T_{i}}%
\frac{L_{T_{i}}}{\rho _{i}},  \label{Pfi}
\end{equation}%
\begin{equation}
P_{b}\equiv \frac{\beta ~v_{T_{i}}}{B_{0}V_{\ast }}=\frac{\beta }{B_{0}}%
\frac{L_{T_{i}}}{\rho _{i}},  \label{Pb}
\end{equation}%
\begin{equation}
P_{c}\equiv \frac{\rho _{i}~\nu }{V_{\ast }}=\frac{L_{T_{i}}}{\lambda _{mfp}}%
,  \label{Pc}
\end{equation}%
where $\lambda _{mfp}=v_{T_{i}}/\nu .$ These parameters measure the
influence that the three stochastic processes (turbulence, RMPs and particle
collisions) have on particle motion. The stochastic functions are
statistically independent, but a strong nonlinear interaction can appear
through particle trajectories due to the space dependence of $\phi $ and $b$.

\section{The semi-analytical solution}

The perpendicular diffusion coefficient is determined as function of $%
P_{\phi },$ $P_{b},$ $P_{c},$\ $\overline{R}$\ and $\tau _{c}.$ The space
dependence of potential makes the transport strongly nonlinear. Nonlinear
effects of the RMPs are expected when $P_{\phi }/P_{b}$ is large. In these
conditions the particles can explore the structure of the stochastic
potential before they leave the correlated zone due to magnetic line
displacements.

A semi-analytical approach based on the decorrelation trajectory method
(DTM, \cite{V1998}) is developed for this multi-stochastic process.

\bigskip

The change of variable from $\mathbf{x}$\ \ to $\mathbf{x}^{\prime }=\mathbf{%
x-}x_{b}(t)\mathbf{e}_{x},$\ where $x_{b}(t)$ is the displacement produced
by the RMPs, permits to define an effective velocity that includes the three
stochastic functions%
\begin{equation}
\mathbf{v}^{eff}(\mathbf{x}^{\prime },t)=-\exp \left( \frac{x^{\prime
}+x_{b}(t)}{\overline{R}}\right) \mathbf{\nabla }^{\prime }\phi (\mathbf{x}%
^{\prime }+x_{b}(t)\mathbf{e}_{x},z(t),t)\times \mathbf{e}_{z}.  \label{vef1}
\end{equation}%
The equations of motion in this frame is%
\begin{equation}
\frac{d\mathbf{x}^{\prime }}{dt}=P_{\phi }\,\mathbf{v}_{eff}(\mathbf{x}%
^{\prime },t)+V_{d}\mathbf{e}_{y}.  \label{1p-eff}
\end{equation}

The transport formally appears in Eq. (\ref{1p-eff}) as produced by a single
stochastic velocity. This is an important simplification, which is effective
if the EC of $\mathbf{v}^{eff}(\mathbf{x}^{\prime },t)$ can be estimated.
The latter is defined as the average over all of the stochastic processes%
\begin{equation}
E_{ij}^{eff}(\mathbf{x}^{\prime },t)=\left\langle \left\langle \left\langle
v_{i}^{eff}(\mathbf{0},0)\ v_{j}^{eff}(\mathbf{x}^{\prime },t)\right\rangle
_{\phi }\right\rangle _{\parallel }\right\rangle _{b}.  \label{ECveff}
\end{equation}

The steps for determining the semi-analytical solution of this transport
problem are presented below. The statistic of the parallel collisional
motion is determined in Subsection 3.1. The probability of the displacements 
$x_{b}(t)$ induced by the RMPs is analyzed in 3.2. The EC of the effective
velocity (\ref{ECveff}) is calculated in Subsection 3.3 and a short review
of the DTM for determining the time dependent diffusion coefficients is
presented in 3.4.

\subsection{Parallel collisional transport}

The first step consists of determining the $z$ component of the trajectories
from Eq. (\ref{1p-z}). This is a well known linear stochastic process that
leads to Gaussian distribution of the trajectories $z(t).$ We give here the
results which are necessary for the following calculations. The probability
that the trajectories are in $z$ at time $t$ is 
\begin{equation}
P_{\parallel }(z,t)\equiv \left\langle \delta \left( z-z(t)\right)
\right\rangle _{\parallel }=\frac{1}{\sqrt{2\pi \left\langle
z^{2}(t)\right\rangle _{\parallel }}}\exp \left( -\frac{z^{2}}{2\left\langle
z^{2}(t)\right\rangle _{\parallel }}\right) ,  \label{ppar}
\end{equation}%
where the mean square displacement (MSD) is%
\begin{equation}
\left\langle z^{2}(t)\right\rangle _{\parallel }=2\int_{0}^{t}d_{\parallel
}(\tau )~d\tau =2P_{c}^{-2}\left[ P_{c}t+\exp (-P_{c}t)-1\right]
\label{zmsd}
\end{equation}%
and 
\begin{equation}
d_{\parallel }(t)=\int_{0}^{t}C_{\parallel }(\tau )~d\tau =P_{c}^{-1}\left[
1-\exp (-P_{c}t)\right] .  \label{dpar}
\end{equation}%
is the time dependent diffusion coefficient along the magnetic field lines.

\subsection{Transport induced by RMPs}

The displacements produced by the RMPs are solutions of 
\begin{equation}
\frac{dx_{b}}{dt}=P_{b}\exp (\frac{x^{\prime }+x_{b}}{\overline{R}}%
)b(y,z)\eta _{\parallel }(t),\ \ \frac{dz}{dt}=\eta _{\parallel }(t).
\label{xb}
\end{equation}%
We consider first a constant confining magnetic field $(\overline{R}%
\rightarrow \infty )$ 
\begin{equation}
\frac{dx_{0}}{dt}=P_{b}b(y,z(t))\eta _{\parallel }(t).  \label{x0}
\end{equation}%
The velocity in the right hand side of this equation is the product of two
stochastic functions, the magnetic field and the collisional velocity. Its
Lagrangian correlation is defined by%
\begin{equation}
C_{v}\equiv \left\langle \left\langle b(0,0)\ b(y,z(t))\right\rangle
_{b}\eta _{\parallel }(0)\eta _{\parallel }(t)\right\rangle _{\parallel }.
\label{cvb}
\end{equation}%
The poloidal confining magnetic field $B_{p}$ leads to the poloidal rotation
of the magnetic lines, which are at the angle $\iota $ with the $z$ axis,
where $tg(\iota )=B_{p}/B_{0}.$ The collisional particle motion along the
magnetic lines $\zeta (t)$ has projections in the toroidal $z(t)=\zeta
(t)\cos (\iota )$ and poloidal $y(t)=\zeta (t)\sin (\iota )$ directions.
Since $\iota $ is small\ $z(t)\cong \zeta (t),$ $y(t)\cong \zeta
(t)B_{p}/B_{0}\cong z(t)B_{p}/B_{0}.$ Using Eq. (\ref{ECmf})\ one obtains%
\begin{equation}
C_{v}\equiv \left\langle \exp \left( -\frac{z^{2}(t)}{2\Lambda _{eff}^{2}}%
\right) \eta _{\parallel }(0)\eta _{\parallel }(t)\right\rangle _{\parallel
},  \label{cvb1}
\end{equation}%
where 
\begin{equation}
\Lambda _{eff}=\frac{\Lambda _{z}}{\sqrt{1+\left( \frac{\Lambda _{z}}{%
\Lambda _{y}}\frac{B_{p}}{B_{0}}\right) ^{2}}}.  \label{Lzef}
\end{equation}%
This shows that the finite $\Lambda _{y}$ of the RMPs determines an
effective correlation length, which is smaller than $\Lambda _{z}.$ The
factor in Eq. (\ref{Lzef}) depends on the number of coils and it is of the
order $\sqrt{1+(m/qn)^{2}},$ where $q$ is the safety factor.

The correlation in Eq. (\ref{cvb1}) is written using $\delta $-function to
impose the condition $z(t)=z$ \ 
\begin{eqnarray}
C_{v} &=&\int_{-\infty }^{\infty }dz~\exp \left( -\frac{z^{2}}{2\Lambda
_{eff}^{2}}\right) \left\langle \delta (z-z(t))~\eta _{\parallel }(0)\eta
_{\parallel }(t)\right\rangle _{\parallel }  \label{cvb2} \\
&=&\frac{1}{2\pi }\int \int_{-\infty }^{\infty }dz~dq\exp \left( -\frac{z^{2}%
}{2\Lambda _{eff}^{2}}+iqz\right) \left\langle \exp (-qz(t))~\eta
_{\parallel }(0)\eta _{\parallel }(t)\right\rangle _{\parallel }  \notag
\end{eqnarray}%
\ \ The average in this equation is%
\begin{eqnarray}
M_{\parallel } &\equiv &\left\langle \exp \left( -iqz(t)\right) ~\eta
_{\parallel }(0)\eta _{\parallel }(t)\right\rangle _{\parallel }  \notag \\
&=&\frac{1}{q^{2}}\partial _{t}\partial _{t_{0}}\left. \left\langle \exp
\left( -iq\int_{t_{0}}^{t}d\tau ~\eta _{\parallel }(\tau )\right)
\right\rangle _{\parallel }\right\vert _{t_{0}=0}.  \label{mpar}
\end{eqnarray}%
Since $\eta _{\parallel }$ and $z(t)$ are Gaussian functions, the average of
the exponential is%
\begin{equation}
\exp \left( -\frac{q^{2}}{2}\int_{t_{0}}^{t}d\tau \int_{t_{0}}^{t}d\tau
^{\prime }~C_{\parallel }(\left\vert \tau -\tau ^{\prime }\right\vert
)\right) .
\end{equation}%
Straightforward calculations lead to%
\begin{equation}
M_{\parallel }=\left( C_{\parallel }(t)-q^{2}d_{\parallel }^{2}(t)\right)
\exp \left( -\frac{q^{2}}{2}\left\langle z^{2}(t)\right\rangle _{\parallel
}\right) .  \label{mpar2}
\end{equation}%
The LVC of the RMP (\ref{cvb2}) becomes after calculating the integrals%
\begin{equation}
C_{v}(t)=P_{b}^{2}\frac{\Lambda _{eff}~\left[ C_{\parallel }(t)(\Lambda
_{eff}^{2}+\left\langle z^{2}(t)\right\rangle _{\parallel })-d_{\parallel
}^{2}(t)\right] }{(\Lambda _{eff}^{2}+\left\langle z^{2}(t)\right\rangle
_{\parallel })^{3/2}}.  \label{cvb3}
\end{equation}%
The time integral of this function gives the time dependent diffusion
coefficient generated by RMPs%
\begin{equation}
D_{b}(t)=P_{b}^{2}\Lambda _{eff}~\frac{d_{\parallel }^{2}(t)}{(\Lambda
_{eff}^{2}+\left\langle z^{2}(t)\right\rangle _{\parallel })^{1/2}}.
\label{drmp}
\end{equation}%
The MSD is obtained by integrating once more 
\begin{equation}
\left\langle x_{0}^{2}(t)\right\rangle _{b\parallel }=2P_{b}^{2}\Lambda
_{eff}~\left( (\Lambda _{eff}^{2}+\left\langle z^{2}(t)\right\rangle
_{\parallel })^{1/2}-\Lambda _{eff}\right)  \label{msdrmp}
\end{equation}

This is the well known \cite{RR} process of subdiffusive transport of the
double diffusion type (magnetic field line diffusion combined with the
collisional particle diffusion along field lines). It corresponds to the
quasilinear transport in stochastic magnetic fields. Nonlinear effects
appear for space dependent magnetic fields with large Kubo numbers \cite%
{VS2015m}.

Any perturbation that lead to the departure of the particles from the field
lines leads to diffusive particle transport. Such a perturbation can be the
collisional perpendicular velocity, plasma rotation or even the small
magnetic drifts determined by the curvature of the magnetic lines. In the
process studied here, the $\mathbf{E}\times \mathbf{B}$ stochastic velocity
produces the strongest decorrelation effect and it has the main role in
determining the RMP diffusion coefficient. However, since the correlation
lengths of $\mathbf{b}$ are very large (at macroscopic scale), the
characteristic time $\tau _{b}$ for the saturation of $D_{b}(t)$ is large.
It is much larger than the Lagrangian characteristic time of the turbulence $%
\tau _{b}>>\tau _{L}.$\ Thus, the RMP transport process can be approximated
by Eq. (\ref{drmp})-(\ref{msdrmp}) during $\tau _{L}.$\ The turbulence and
other decorrelation mechanisms have negligible effects at such small times.

We note that the complex structure of the magnetic field lines generated by
the RMPs (see for instance \cite{review}) does not influence the transport
process studied here. The magnetic structure with stochastic regions and
island chains is evidenced by following the magnetic lines for many toroidal
rotations. Or, the turbulence has a parallel correlation length of the order
of $\overline{R},$ which means that after one toroidal rotation the
particles leave the magnetic lines. In other words, the particles do not
"see" the complex structure of the magnetic lines in the presence of
turbulence.

We evaluate now the effect of the gradient of the confining magnetic field
taking into account the $\overline{R}$ dependent factor in Eq. (\ref{xb}).
It can be written as%
\begin{equation}
\frac{dx_{b}}{dt}=P_{b}\exp (\frac{x^{\prime }+x_{b}}{\overline{R}})\frac{%
dx_{0}}{dt}.  \label{xb1}
\end{equation}%
The solution in terms of $x_{0}(t)$\ is%
\begin{eqnarray}
x_{b}(t) &=&-\overline{R}~\ln \left( 1-\frac{x_{0}(t)}{\overline{R}}\exp
\left( \frac{x^{\prime }}{\overline{R}}\right) \right)  \label{sol} \\
&\cong &x_{0}(t)\exp \left( \frac{x^{\prime }}{\overline{R}}\right) +\frac{%
x_{0}^{2}(t)}{2\overline{R}}\exp \left( \frac{2x^{\prime }}{\overline{R}}%
\right) +\frac{x_{0}^{3}(t)}{3\overline{R}^{2}}\exp \left( \frac{3x^{\prime }%
}{\overline{R}}\right) +...  \notag
\end{eqnarray}%
The average of this equation is%
\begin{equation}
\left\langle x_{b}(t)\right\rangle \cong \frac{\left\langle
x_{0}^{2}(t)\right\rangle }{2\overline{R}}\exp \left( \frac{2x^{\prime }}{%
\overline{R}}\right) \cong \frac{\left\langle x_{0}^{2}(t)\right\rangle }{2%
\overline{R}},  \label{xbmed}
\end{equation}%
and the average of the square is%
\begin{equation}
\left\langle x_{b}^{2}(t)\right\rangle \cong \left\langle
x_{0}^{2}(t)\right\rangle \exp \left( \frac{2x^{\prime }}{\overline{R}}%
\right) ,  \label{xb2}
\end{equation}%
where the linear approximation in the small parameter $1/\overline{R}$ was
taken.\ 

The probability of the RMP generated displacements is Gaussian in the linear
approximation in $1/\overline{R}$ 
\begin{equation}
P_{b}(x_{b},t)\cong \frac{1}{\sqrt{2\pi \left\langle
x_{b}^{2}(t)\right\rangle }}\exp \left( -\frac{\left( x_{b}-\left\langle
x_{b}(t)\right\rangle \right) ^{2}}{2\left\langle x_{b}^{2}(t)\right\rangle }%
\right)  \label{Prmp}
\end{equation}%
\ \ \ \ 

Thus, the gradient of the confining magnetic field generates an average
displacement that is proportional to the MSD produced by the RMPs, and to
the gradient. Such average displacement generates an average velocity
(direct transport) 
\begin{equation}
V_{b}(t)\equiv \frac{d\left\langle x_{b}(t)\right\rangle }{dt}=\dfrac{%
D_{b}(t)}{\overline{R}}~  \label{Vb}
\end{equation}%
The velocity $V_{b}$ is positive, directed against the gradient of the
magnetic field, towards plasma boundary. It is transitory for the
subdiffusive transport, and a finite asymptotic value exists only in the
presence of a process of decorrelation of the particles from the field
lines. We are interested here in the nonlinear effects produced by the
average displacement $\left\langle x_{b}(t)\right\rangle $ on the turbulent
transport. \ 

\subsection{The EC of the effective velocity}

The averages over the three stochastic functions that appear in the
effective velocity are calculated according to the definition (\ref{ECveff}).

The average over the stochastic potential yields the EC of the $\mathbf{E}%
\times \mathbf{B}$ drift components (\ref{vEC})%
\begin{eqnarray}
M_{ij} &\equiv &\left\langle v_{i}^{eff}(\mathbf{0},0)\ v_{i}^{eff}(\mathbf{x%
}^{\prime },t)\right\rangle _{\phi }  \label{mfi} \\
&=&C_{ij}(\mathbf{x}^{\prime }+x_{b}(t)\mathbf{e}_{x},z(t),t)~\exp \left( 
\frac{x^{\prime }+x_{b}(t)}{\overline{R}}\right) .  \notag
\end{eqnarray}

The average over the parallel collisional velocity $\eta _{\parallel }(t)$
is obtained using the probability (\ref{ppar}) of $z(t)$. It applies in the
case of the EC (\ref{rEC}) only to the $z$\ dependent factor, which becomes%
\begin{eqnarray}
\left\langle \exp \left( -\frac{z^{2}}{2\lambda _{z}^{2}}\right)
\right\rangle _{\parallel } &=&\int_{-\infty }^{\infty }\exp \left( -\frac{%
z^{2}}{2\lambda _{z}^{2}}\right) P_{\parallel }(z,t)dz  \notag \\
&=&\frac{\lambda _{z}}{\sqrt{\lambda _{z}^{2}+\left\langle
z^{2}(t)\right\rangle _{\parallel }}}.  \label{mz}
\end{eqnarray}%
\ The average over the RMPs is calculated using the probability (\ref{Prmp})
of the magnetic displacements. This average changes only the radial factor
in the EC (\ref{rEC}) multiplied by the gradient of the confining magnetic
field, which is%
\begin{eqnarray*}
M_{b} &\equiv &\left\langle \exp \left( -\frac{(x^{\prime }+x_{b}(t))^{2}}{%
2\lambda _{x}^{2}}\right) \exp \left( \frac{x^{\prime }+x_{b}(t)}{\overline{R%
}}\right) \right\rangle _{b} \\
&=&\int_{-\infty }^{\infty }\exp \left( -\frac{(x+x_{b})^{2}}{2\lambda
_{x}^{2}}+\frac{x^{\prime }+x_{b}}{\overline{R}}\right) P_{b}(x_{b},t)dx_{b}
\end{eqnarray*}%
One obtains after neglecting a small term of second order in $1/\overline{R}$%
\begin{eqnarray}
M_{b} &=&\frac{\lambda _{x}}{\sqrt{\lambda _{x}^{2}+\left\langle
x_{b}^{2}(t)\right\rangle }}\exp \left[ -\frac{(x^{\prime }+\left\langle
x_{b}(t)\right\rangle )^{2}}{2(\lambda _{x}^{2}+\left\langle
x_{b}^{2}(t)\right\rangle )}\right]  \notag \\
&&\exp \left( \frac{x^{\prime }+\left\langle x_{b}(t)\right\rangle }{%
\overline{R}}\frac{\lambda _{x}^{2}}{\lambda _{x}^{2}+\left\langle
x_{b}^{2}(t)\right\rangle }\right) .  \label{mb}
\end{eqnarray}

Finally, the EC of the effective velocity (\ref{ECveff}) can be written as

\begin{equation}
E_{ij}^{eff}(\mathbf{x}^{\prime },t)=-\varepsilon _{ik}\varepsilon
_{jl}\partial _{k}\partial _{l}\left[ E^{eff}(\mathbf{x}^{\prime },t)\right]
\exp \left( \frac{x^{\prime }+\left\langle x_{b}(t)\right\rangle }{R^{eff}}%
\right) ,  \label{ECvef1}
\end{equation}%
where%
\begin{equation}
E^{eff}(\mathbf{x}^{\prime },t)=\frac{\lambda _{x}}{\lambda _{x}^{eff}}\frac{%
\lambda _{z}}{\lambda _{z}^{eff}}\exp \left( -\frac{(x^{\prime
}+\left\langle x_{b}(t)\right\rangle )^{2}}{2(\lambda _{x}^{eff})^{2}}%
\right) \partial _{y}\left( \exp \left( -\frac{y^{2}}{2\lambda _{y}^{2}}%
\right) y\right) ,  \label{ECeffp}
\end{equation}%
\begin{equation}
\lambda _{x}^{eff}(t)=\sqrt{\lambda _{x}^{2}+\left\langle
x_{b}^{2}(t)\right\rangle },  \label{lxef}
\end{equation}%
\begin{equation}
R^{eff}=\overline{R}\frac{\left( \lambda _{x}^{eff}\right) ^{2}}{\lambda
_{x}^{2}},  \label{Reff}
\end{equation}%
\begin{equation}
\lambda _{z}^{eff}(t)=\sqrt{\lambda _{z}^{2}+\left\langle
z^{2}(t)\right\rangle _{\parallel }}.  \label{lzef}
\end{equation}%
The last factor in Eq. (\ref{ECvef1}) is determined by the gradient of the
confining magnetic field, and the other have the same structure as in Eq. (%
\ref{vEC}), which relates the correlation of the drift velocity to the EC of
the potential. An effective potential with the EC (\ref{ECeffp}) can be
defined for $\mathbf{v}^{eff}(\mathbf{x}^{\prime },t).$

The difference between the effective velocity and the $\mathbf{E}\times 
\mathbf{B}$ drift velocity consists of the change of the EC of the
potential. Comparing Eq. (\ref{ECeffp}) with Eq. (\ref{rEC}), one can see
that the modification concern the radial dependence of the EC. It consists
of the shift of the maximum with the average displacement generated by the
RMPs and of the increase of the correlation length, which becomes a
time-dependent function (\ref{lxef}). The RMPs also determine the decay in
time of the amplitude as $\lambda _{x}/\lambda _{x}^{eff}(t).$

The parallel motion eliminates the $z$ dependent factor in (\ref{rEC}) and
leads to the time decay of the amplitude of the effective potential as $%
\lambda _{z}/\lambda _{z}^{eff}(t).$ the exponential decay with $z$ is
transformed into a slow decorrelation (the decay of the amplitude as $1/%
\sqrt{t}).$\ The parallel collisional velocity also determines the
subdiffusive behaviour of the MSD of the RMP displacements $x_{b}(t).$ This
effect leads to a slow decorrelation of $\mathbf{v}^{eff}$ in the radial
direction, much slower than in the absence of parallel collisions when the
RMP radial transport is diffusive. \ \ 

The factor determined by the gradient of the toroidal magnetic field in Eq. (%
\ref{ECvef1}) is modified by the RMPs, which determine a shift of the
maximum and the increase of the gradient length (\ref{Reff}). This means
that the effect of $\overline{R}$\ decreases in time.\ \ 

\subsection{The DTM}

\ The multi-stochastic process that describes turbulent transport in the
presence of RMPs was reduced to the problem of 2-dimensional transport in an
effective velocity field that has the EC (\ref{ECvef1}) with $E^{eff}(%
\mathbf{x}^{\prime },t)$ given by (\ref{ECeffp}).

We use the DTM \cite{V1998} for determining the time dependent diffusion
coefficient. This method is based on a set of deterministic trajectories,
the decorrelation trajectories (DTs), which are obtained from the EC of the
effective velocity. We define a set of subensembles $S$ with given values of
the stochastic functions at the origin: 
\begin{equation}
\phi ^{eff}(\mathbf{0},0)=\phi ^{0},\quad \mathbf{v}^{eff}(\mathbf{0},0)=%
\mathbf{v}^{0}.  \label{S}
\end{equation}%
The effective velocity is in each subensemble S is a Gaussian field with the
average 
\begin{equation}
V_{i}^{S}(\mathbf{x},t)=\phi ^{0}E_{\phi i}^{eff}(\mathbf{x}^{\prime
},t)+v_{1}^{0}E_{1i}^{eff}(\mathbf{x}^{\prime },t)+v_{2}^{0}E_{2i}^{eff}(%
\mathbf{x}^{\prime },t),  \label{VS}
\end{equation}%
where $E_{\phi i}^{eff}(\mathbf{x}^{\prime },t)=-\varepsilon _{ik}\partial
_{k}E^{eff}(\mathbf{x}^{\prime },t)$ are the correlations of the potential
with the effective velocity. The DTs are approximate average trajectories in
the subensembles obtained by solving the equation 
\begin{equation}
\frac{d\mathbf{X}^{S}(t)}{dt}=P_{\phi }\mathbf{V}^{S}(\mathbf{X}%
^{S}(t),t)\exp \left( \frac{X^{S}(t)+\left\langle x_{b}(t)\right\rangle }{%
R^{eff}}\right) +V_{d}\mathbf{e}_{y}.  \label{DT}
\end{equation}%
The fluctuations of the trajectories are neglected in this equation. This
approximation is supported by the high degree of similarity of the
trajectories in a subensemble, which is determined by the supplementary
initial conditions (\ref{S}), and by the small small amplitude of the
velocity fluctuations in a subensemble \cite{VS2004}.

The time dependent diffusion coefficient and the average radial displacement
are obtained by summing the contributions of all subensembles (see \cite%
{V1998} for details) 
\begin{equation}
D_{i}(t)=\int d\phi ^{0}d\mathbf{v}^{0}P(\phi ^{0})P(\mathbf{v}%
^{0})~v_{i}^{0}X_{i}^{S}(t),  \label{Ddt}
\end{equation}%
\begin{equation}
\left\langle x(t)\right\rangle =\int d\phi ^{0}d\mathbf{v}^{0}P(\phi ^{0})P(%
\mathbf{v}^{0})~X_{i}^{S}(t).  \label{xmed}
\end{equation}

The direct contribution of the transport produced by the RMPs has to be
added. Thus the diffusion coefficients in physical units are

\begin{equation}
D_{i}^{tot}(t;P_{\phi },P_{b},P_{c},\overline{R})=\frac{\Phi }{B_{0}}\left(
D_{i}(t;P_{\phi },P_{b},P_{c},\overline{R})+D_{b}(t;P_{b},P_{c},\overline{R}%
)\right)  \label{Dtot}
\end{equation}%
The asymptotic diffusion coefficients are

\begin{equation}
\chi _{i}^{tot}(P_{\phi },P_{b},P_{c},\overline{R})=\frac{\Phi }{B_{0}}%
\left( \chi _{i}(P_{\phi },P_{b},P_{c},\overline{R})+\chi _{b}(P_{b},P_{c},%
\overline{R})\right) ,  \label{dast}
\end{equation}%
\ 

\bigskip

Thus, the diffusion coefficients $D_{i}^{b}(t)$ are obtained from Eq. (\ref%
{Ddt}) using the solutions of Eq. (\ref{DT}) for the DTs. The latter have to
be numerically calculated, although they are very simple. A computer code
was developed for the calculation\ of the decorrelation trajectories, of the
running diffusion coefficient (\ref{Ddt}) and of the average displacement (%
\ref{xmed}). The numerical calculations are at the microcomputer level with
runs of the order of few minutes.

\section{The effects of RMPs on turbulent transport}

We analyze here the effects of the RMPs on transport as function of
turbulence parameters. The effects produced by the increase of RMP intensity
on turbulence are not considered. The aim is to identify the direct change
of transport.

The multi-stochastic process that describes the turbulent transport in the
presence of RMPs depends on twelve physical parameters. This number is
reduce to 10 using dimensionless variables. This large number of parameters
imposes a first analysis of their ranges and of the importance of each term
before quantitative evaluations.

The main parameters are $P_{\phi }$ (\ref{Pfi}) and $P_{b}$\ (\ref{Pb}), the
amplitudes of the turbulence and of the RMPs, respectively. We consider ion
temperature (ITG) driven turbulence that has correlation lengths of the
order of $\rho _{i}.$ The values taken in the figures are $\overline{\lambda 
}_{x}\equiv \lambda _{x}/\rho _{i}=4$ and $\overline{\lambda }_{y}\equiv
\lambda _{y}/\rho _{i}=2.$\ The parallel correlation length $\lambda _{z}$
is of the order of $R,$\ which leads to $\overline{\lambda }_{z}\equiv
\lambda _{z}/L_{T_{i}}=\overline{R}/L_{T_{i}},$ which is of the order of the
ITG parameter. A value that corresponds to well developed ITG turbulence is
taken for the calculations in the figures ($\overline{\lambda }_{z}=6$ ).
The drift velocity $V_{d}$ is of the order of the diamagnetic velocity,
which leads to $\overline{V}_{d}\lesssim 1.$ Thus, the analysis of the
dependence of the transport on four of the parameters of the model\ is not
necessary since they have narrow variation range and determine weak
effects.\ Also, the collision parameter $P_{c}$ (\ref{Pc}) has a weaker
effect. It essentially separates the ballistic and the diffusive behavior of
the parallel trajectories and influences the MSD of the displacements
produced by the RMPs.

The normalized amplitude of the RMPs can be written as
\begin{subequations}
\begin{equation}
P_{b}=(\beta /B_{0})(L_{T_{i}}/R)(R/a)(a/\rho _{i})\simeq 0.5(\beta
/B_{0})(a/\rho _{i}).  \label{Pbval}
\end{equation}
It depends on the amplitude of the magnetic field and on plasma size. Its
values are $P_{b}\lesssim 1$ in the actual experimental conditions and will
increase to $P_{b}\simeq 10$ for ITER. The effects of the RMPs on turbulent
transport are analyzed for a large range of amplitudes $\left[ 0.01,~100%
\right] .$

We begin by a short presentation of the trapping process and of the
transport regimes obtained in a turbulence with the EC of the type (\ref{rEC}%
) in the absence of the RMPs (Subsection 4.1). The effects of the RMPs on
the transport and on the turbulent pinch determined by the gradient of the
toroidal magnetic field are discussed in 4.2.

\subsection{Trajectory trapping \ \ \ \ \ }

Particle trajectories in turbulent plasmas can have both random and
quasi-coherent aspects. A typical trajectory is a random sequence of long
jumps and trapping events that consists of winding on almost closed paths.
Trapping introduces quasi-coherent aspects in trajectory statistics. It
determines a large degree of coherence in the sense that bundles of
trajectories that start from neighboring points remain close for very long
time compared to the eddying time \cite{VS2004}. This process generates
intermittent, quasi-coherent structures of trajectories similar to fluid
vortices.

A strong interdependence exists between ion trajectory statistics and the
evolution of the drift type turbulence. This interaction is rather complex
and it involves both the random and the quasi-coherent aspects, but with
completely different effects. Trajectory diffusion has a stabilizing effect
on turbulence \cite{VSRJP2015} while trajectory trapping leads to strong
nonlinear effects \cite{Vlad2013}. The strength of each of these processes
depends on the stage in the evolution of turbulence. The transport is
related to the stochastic aspect of trajectories, and trajectory structures
have a hindering effect. The trapped particle do not contribute to
transport, but they represent a reservoir of transport. Any perturbation
that liberates particles leads to increased transport and to anomalous
transport regimes.

The turbulent transport in the absence of RMPs ($P_{b}=0)$ was studied in 
\cite{VS2013} for a potential with EC of the type (\ref{rEC}). We present
here a short review of these results and of their physical image.

The process is nonlinear due to the $\mathbf{x}$ dependence of the
stochastic potential. The nonlinearity manifests as trajectory trapping or
eddying due to the Hamiltonian structure of Eqs. (\ref{1p-x}-\ref{1p-y}),
which lead\ to the invariance of the Lagrangian potential for $\tau
_{c}\rightarrow \infty ,$\ $\lambda _{z}\rightarrow \infty $ and $V_{d}=0.$
The trajectories remain on the contour lines of the potential, and the
transport is subdiffusive in these conditions. The time variation of the
potential and/or particle parallel motion when $\lambda _{z}$\ is finite
represent decorrelation mechanisms, which lead to finite asymptotic values
of the diffusion coefficient. Depending on the strength of these
perturbations, represented by decorrelation characteristic times $\tau _{d}$%
, trajectory trapping is partially or completely eliminated. The condition
for the existence of trapped trajectories is $\tau _{d}>\tau _{fl},$\ where $%
\tau _{d}\equiv \tau _{c}\tau _{z}/(\tau _{c}+\tau _{z}),$ $\tau _{z}$\ is
the parallel decorrelation time, and $\tau _{fl}=\lambda _{x}/V_{x}+\lambda
_{y}/V_{y}$\ is the time of flight of the particles or the eddying time. The
average velocity $V_{d}$ also influences the trapping, but in a different
way. It determines an average potential $xV_{d}$\ that adds to the
stochastic potential $\phi (\mathbf{x}).$\ The total potential has a
strongly modified structure. Bunches of open contour lines between islands
of closed lines appear for small $V_{d}$ ($V_{d}<V,$ where $V=\sqrt{%
V_{x}^{2}+V_{y}^{2}}$ is the amplitude of the stochastic velocity). As $%
V_{d} $ increases, the surface occupied by the islands of closed lines
decreases and vanishes for $V_{d}>V.$ Thus, the average velocity $V_{d}$\
eliminates trajectory trapping, but not through a decorrelation mechanism.

The conditions for the existence of trajectory trapping are $\tau _{d}>\tau
_{fl}$ and $V_{d}<V.$\ In terms of the dimensionless parameters used in this
paper, these conditions are $\overline{\tau }_{d}\equiv \tau _{d}/\tau _{0}>%
\overline{\lambda }_{x}\overline{\lambda }_{y}/P_{\phi }$ and $\overline{V}%
_{d}\overline{\lambda }_{x}<P_{\phi }$.\ 

In the absence of trapping the transport is quasilinear. At small
decorrelation time ($\overline{\tau }_{d}P_{\phi }<\overline{\lambda }_{x}%
\overline{\lambda }_{y}),$ $\chi _{x}=V_{x}^{2}\tau _{d},$\ and it does not
depend on $V_{d}.$ When $\tau _{d}$\ is large and the amplitude of the
turbulence is smaller than $V_{d}$\ ($P_{\phi }<\overline{V}_{d}\overline{%
\lambda }_{x}$), the diffusion coefficient decreases with the increase of $%
\tau _{d}$\ as $\chi _{x}=(\phi /B_{0}V_{d}^{2})^{2}/\tau _{d}.$ A different
scaling in the parameters of the turbulence is obtained in the presence of
trapped trajectories, but the decay with $\tau _{d}$ persists. In these
conditions, $\chi _{x}$ depends on the correlation lengths of the turbulence
and on the shape of the EC\ \cite{VS2015}.\ \ \ 

\subsection{RMP effects on transport}

First, we analyze the time dependent diffusion coefficient in the absence of
decorrelation ($\tau _{c}\rightarrow \infty ,$ $\lambda _{z}\rightarrow
\infty )$ in order to identify the main effects of the RMPs.


\begin{figure}[tbp]
\centerline{\includegraphics[height=8cm]{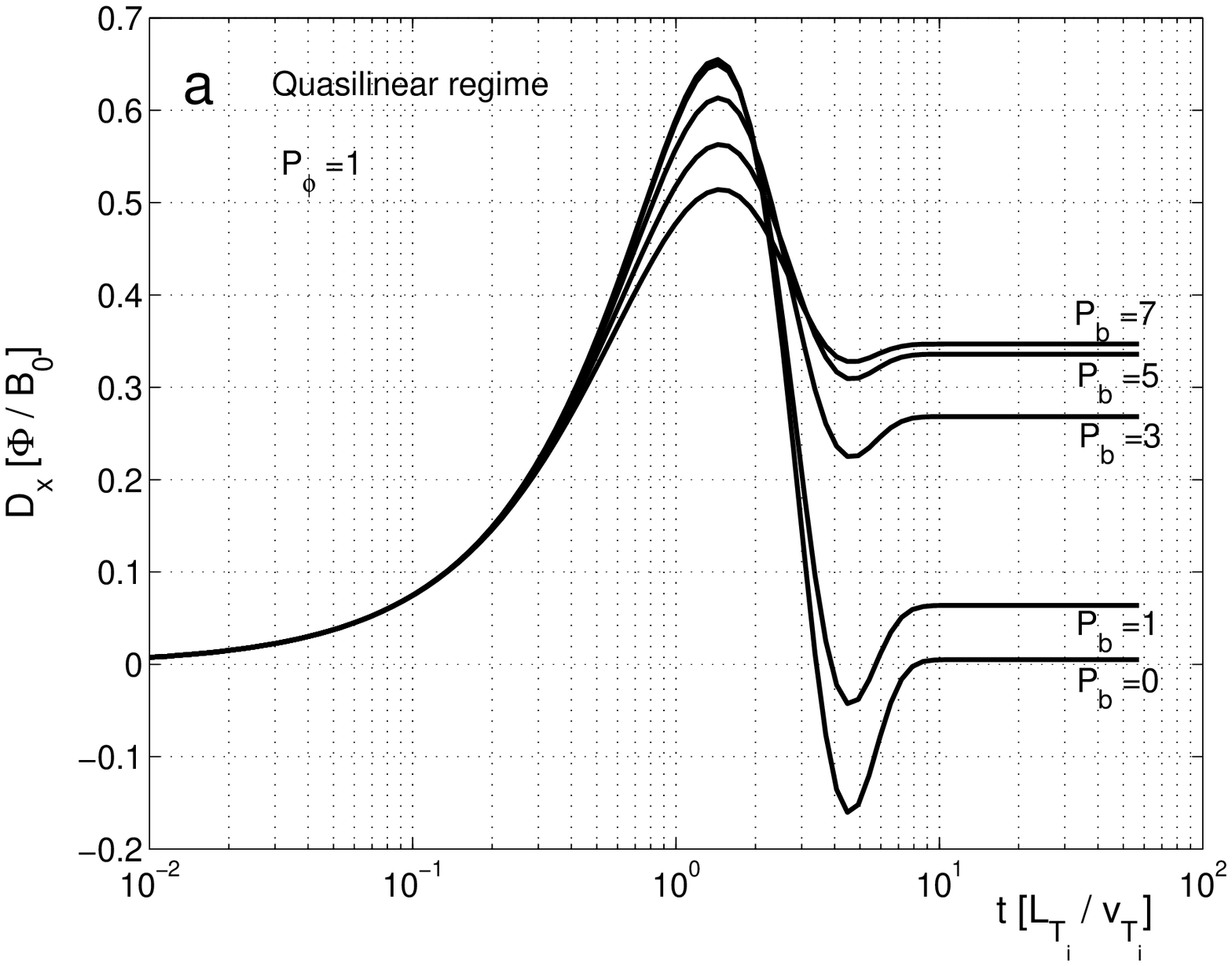}} \centerline{%
\includegraphics[height=8cm]{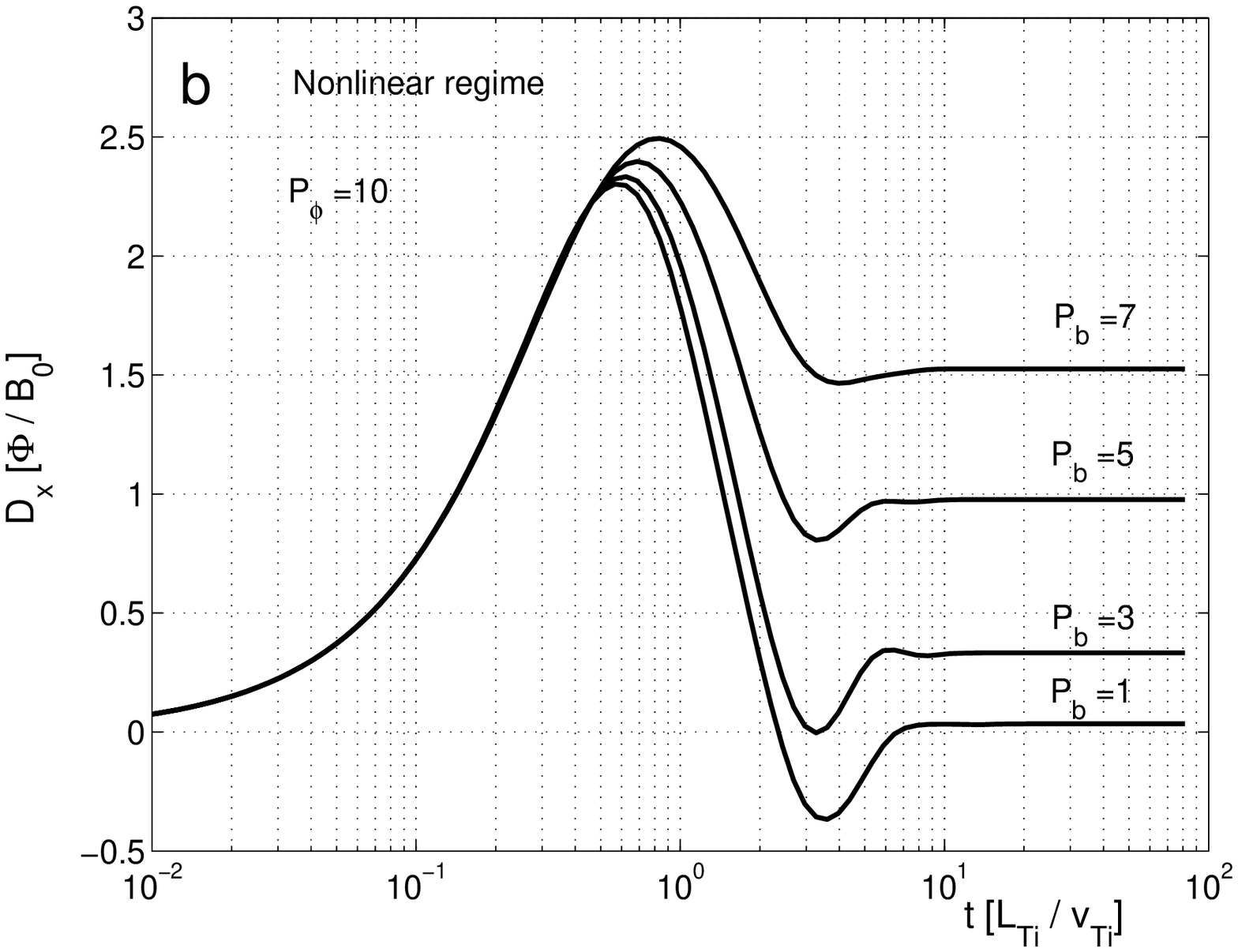}}
\caption{Time dependent diffusion coefficient for $\protect\tau%
_{c}\rightarrow \infty ,$ $\protect\lambda _{z}\rightarrow \infty, $ and the
values of the RMP amplitude that label the curves (a) in the quasilinear
regime with $P_\protect\phi=1$, and (b) in the nonlinear regime with $P_%
\protect\phi=10$. $\overline{V}_d=1$.}
\label{Figure1}
\end{figure}


Typical examples of the time dependent diffusion coefficient in the presence
of RMPs are shown in Figure 1.a. for the quasilinear conditions ($P_{\phi
}=1),$ and in Figure 1.b. for the nonlinear case $(P_{\phi }=10).$ One can
see that the RMPs determine the saturation of $D_{x}(t)$ at finite values in
all cases.

In the absence of the RMPs, the transport is subdiffusive in both
quasilinear and nonlinear regimes. The RMPs make the transport diffusive,
which means that they provide a decorrelation mechanism. Therefore, the RMPs
enhance the stochastic aspects of the trajectories by destroying the
quasi-coherent structures. The increase of the transport coefficient is
expected in such conditions.

Since the RMPs provide a decorrelation mechanism, their effects on the
asymptotic diffusion coefficients should be understood from the analysis of
the competition with the other decorrelation mechanisms.

The RMPs could produce decorrelation by the average velocity (\ref{Vb}) or
by trajectory spreading (\ref{xb2}) since both processes are induced by
RMPs. The examination of the EC of the effective potential (\ref{ECeffp})
shows that the average displacement $\left\langle x_{b}(t)\right\rangle $
does not modify the EC, but it only determines a shift of the EC. The RMP
average velocity does not contribute to the decorrelation. It actually
determines a radial drift of the stochastic potential.

The asymptotic diffusion coefficient $\chi _{x}$ is shown in Figure 2 as a
function of the normalized amplitude of the RMPs for the quasilinear regime
(continuous line) and for the nonlinear regime (dashed line). One can see
that the transport both regimes is not affected at small $P_{b}$, and that
there is a smooth transition to a rather strong degradation of the
confinement. At larger amplitudes, the tendency is reversed and the
diffusion coefficient decreases.

Similar dependences on $P_{b}$ are found in the quasilinear and nonlinear
regimes. There are however some important differences. The maximum diffusion 
$\chi _{x}^{\max }$ and the corresponding RMP amplitude $P_{b}^{\max }$ do
not depend on turbulence amplitude $P_{\phi }$ in the nonlinear transport
while they are increasing functions of $P_{\phi }$ in the quasilinear
regime. The increase of $\chi _{x}$ in the quasilinear regime is much
smaller than in the nonlinear regime. The maximum amplification factor is
four times larger in the nonlinear regime in the examples shown in Figure 2.
Also the maximum corresponds to smaller amplitudes of the RMPs in the
quasilinear case.

The nonlinear dependence of $\chi _{x}$ on $P_{b}$ seen in Figure 2 is
explained by the decorrelation effect produced by the RMPs. The
decorrelation time $\tau _{b}$ is determined as solution of
\end{subequations}
\begin{equation}
\left\langle x_{b}^{2}(\tau _{b})\right\rangle =\overline{\lambda }_{x}^{2},
\label{taudec}
\end{equation}%
\ where $\left\langle x_{b}^{2}(t)\right\rangle $\ is given by Eq. (\ref{xb2}%
). Since $\left\langle x_{b}^{2}(t)\right\rangle \sim P_{b}^{2},$ $\tau _{b}$
decreases when $P_{b}$\ increases.\ The decorrelation of the turbulence is
determined by the parallel motion, and the corresponding decorrelation time $%
\tau _{z}$\ is the solution of%
\begin{equation}
\left\langle z^{2}(\tau _{z})\right\rangle =\overline{\lambda }_{z}^{2}.
\label{taupar}
\end{equation}%
\ One obtains $\tau _{z}=$\ $\overline{\lambda }_{z}$ if $\tau _{z}\ll
1/P_{c}$\ (or $\overline{\lambda }_{z}P_{c}\ll 1),$\ and $\tau _{z}=$\ $%
\overline{\lambda }_{z}^{2}P_{c}$\ if $\tau _{z}\gg 1/P_{c}$ (or $\overline{%
\lambda }_{z}P_{c}\gg 1).$\ At small $P_{b},$\ $\tau _{b}$\ is large ($\tau
_{z}\ll \tau _{b}),$\ which means that the decorrelation is determined by
the parallel motion and the RMPs do not influence the diffusion process. As
the RMP amplitude increases, $\tau _{b}$\ \ decreases\ and the RMP
decorrelation become dominant. They determine the release of an increasing
fraction of trapped trajectories, which contributes to the diffusion and
increases $\chi _{x}.$\ This tendency is reversed after the release of all
trajectories. In these conditions, $\chi _{x}=V_{x}^{2}\tau _{b},$\ and it
decreases with the increase of $P_{b}.$

Typical values of $P_{b}$\ in the present experiments are less than one. In
this range, the quasilinear regime is characterized by a stronger influence
of the RMPs than the nonlinear transport (Figure 2). The smooth threshold in
the dependence of the diffusion coefficient on the amplitude of the RMPs is
in agreement with the experiments \cite{turb-L}, \cite{turb4}.


\begin{figure}[tbp]
\centerline{\includegraphics[height=8cm]{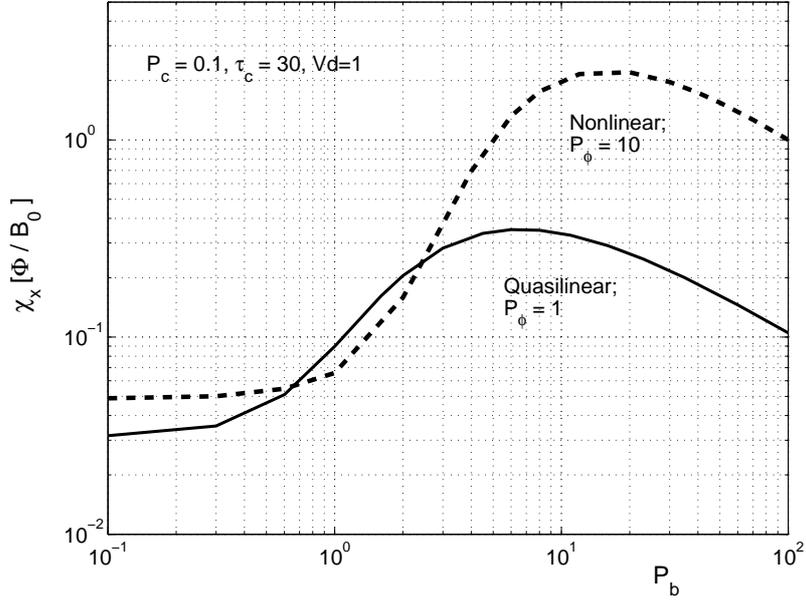}}
\caption{Asymptotic diffusion coefficient as a function of the normalized
amplitude of the RMPs for a set of parameters that corresponds to the
nonlinear regime.}
\label{Figure2}
\end{figure}


\bigskip


\begin{figure}[tbp]
\centerline{\includegraphics[height=8cm]{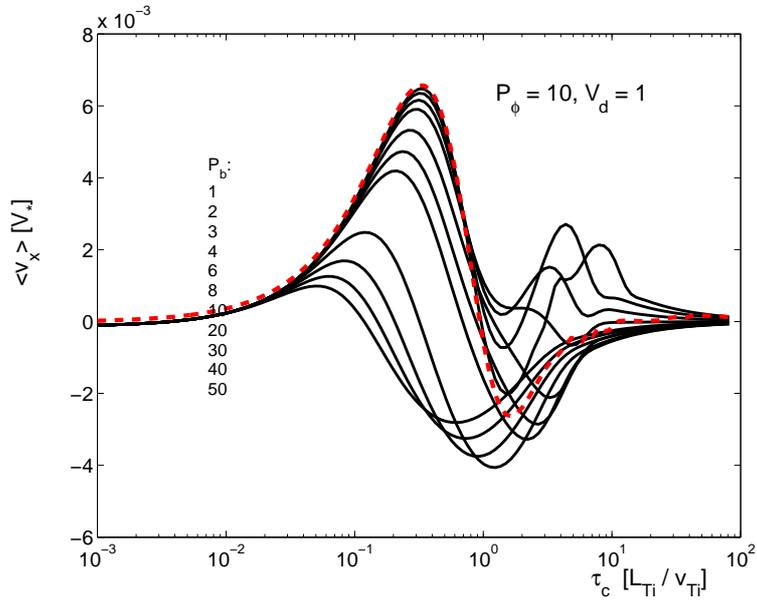}}
\caption{The average velocity as function of the correlation time for a
large domain of the amplitudes of the RMPs, $P_b=1,...,50$. The dashed (red)
curve is the average velocity in the absence of RMPs.}
\label{Figure3}
\end{figure}


The gradient of the toroidal magnetic field generates an average velocity (%
\ref{Vb}) of the trajectories $x_{b}(t)$ determined by the RMPs. A similar
effect was found in the case of turbulent plasmas \cite{V2006}, \cite{V2008}%
. The turbulent pinch velocity is positive in the quasilinear regime
corresponding to small $\tau _{c},$\ and it becomes negative (directed
inward) in the nonlinear regime (large $\tau _{c})$ (see Figure 3, the
dashed line).\ 

We determine here the influence of the RMPs on the turbulent pinch velocity.
As seen in Figure 3, the effect is complex and it depends on the amplitude
of the RMPs, and on the decorrelation time of the turbulence. The RMPs lead
to continuous decrease of the quasilinear pinch. In the nonlinear regime, a
much more complicated dependence on $P_{b}$ is found. A fast growth of $%
\left\langle v_{x}\right\rangle $ appears at small $P_{b},$\ which reaches
positive (outward) values. After a maximum for $P_{b}\simeq 1,$\ the pinch
decreases and becomes negative. Its minimum decreases and moves toward
smaller $\tau _{c}$ as $P_{b}$ increases. The maximum negative velocity is
found at $P_{b}\simeq 20$ and $\tau _{c}\simeq 1.$ At very large values of
the order $P_{b}>20,$ the absolute value of the minimum $\left\langle
v_{x}\right\rangle $ decreases.

The effect of the pinch velocity is determined by the dimensionless
parameter $p\equiv L_{T_{i}}\left\langle v_{x}\right\rangle /\chi _{x},$ the
peaking factor. It is the estimation of the ratio of the average and the
diffusive displacements. The peaking factor for the direct contribution of
the RMPs is small $p_{RMP}=L_{T_{i}}/\overline{R}\cong 0.2.$ The turbulent
peaking factor decreases due to the RMPs, because the diffusion coefficient
increases. However, it can be much larger than $p_{RMP}.$ Values of the
order $\gtrsim 1$\ can be attained\ only for $P_{b}\simeq 1$\ and for large
size plasmas with $a/\rho _{i}\simeq 1000.$ As seen in Figure 3, the pinch
is positive at such values, and it contributes to confinement degradation.

\section{Conclusions and discussions}

The direct effects of the RMPs on turbulent transport were analyzed. The
diffusion coefficient and the pinch velocity were determined as functions of
the turbulence parameters and of the RMPs amplitude $P_{b}$ in the framework
of the test particle approach using a semi-analytical method, the DTM. We
underline that the effects of the RMPs on turbulence are neglected in this
evaluations. The influence of the stochastic magnetic field generated by the
RMPs are rather complex, especially in the nonlinear regime that corresponds
to trajectory trapping or eddying. One of the effects, which is well
demonstrated and understood, is the attenuation of the modes determined by
the increased diffusion, which leads to the decrease of $\Phi $. But other
processes could have opposite effects on turbulence amplitude or they can
even generate a different type of turbulence.

We have shown that the effects observed in experiments (increased turbulent
transport and generation of outward pinch) occur even when the turbulence is
not modified by the RMPs. A direct influence of the RMPs on transport is
produced through a decorrelation mechanism.

The dependence of the diffusion coefficient on the amplitude of the RMPs is
nonlinear (Figure 2). A smooth threshold exists\ at small\ $P_{b}.$\ It is
determined by the condition that the characteristic time of the RMP
decorrelation that is a decreasing function on $P_{b}$\ should be smaller
than the parallel decorrelation time. At larger $P_{b},$\ the increase of
the transport\ coefficients appear in both quasilinear and nonlinear
regimes, with stronger effect in the first case. The increase is limited,
and, after a maximum, the transport enters into a decaying regime. The
maximum is very large in the nonlinear regime ($40$ times larger than in the
absence of the RMPs)\ and it appears at very large amplitudes ($P_{b}\simeq
20).$ In the quasilinear regime, the the maximum is much smaller (by a
factor five in the example in Figure 2) and the RMP amplitude at the maximum
is of the order $P_{b}^{\max }\simeq 5.$ Both $P_{b}^{\max }$ and $\chi
_{x}^{\max }$ decrease as the turbulence amplitude decreases.

These results are in agreement with the experiments, which correspond to
values of the RMP amplitude $P_{b}\lesssim 1$ and lead to increases of the
diffusion coefficients of the order $50-100\%.$ These values of $P_{b}$\ are
close to the smooth threshold where similar variation of $\chi _{x}$ can be
seen in Figure 2. According to our model, the results of the present
experiments cannot be extrapolated to ITER conditions. A much faster
increase with $P_{b}$\ occurs at larger $P_{b}.$ We have found a large
difference between the nonlinear and the quasilinear transport at $%
P_{b}\simeq 10,$ which corresponds to the RMPs of the the order of $1\%$ in
ITER plasmas. The confirmation of the transition of the ITG transport from
the Bohm to the gyro-Bohm regime and demonstration that the gyro-Bohm
transport is of quasilinear type are very important in this context.

We have also analyzed the effect of the RMPs on the turbulent pinch
velocity. We have shown that at large correlation times, the negative
(inward) drift (dashed curve in Figure 3) is reduced by the RMPs, then they
generate a positive (outward) drift that is maximum for $P_{b}\simeq 1.$ At
larger $P_{b}$ the average velocity decreases and becomes again negative,
which correspond to the prediction of an inward pinch in ITER conditions,
but with small values of the peaking.

\bigskip

\bigskip

\textbf{Acknowledgements}

This work was supported by the Romanian Ministry of National Education under
the contract 1EU-10 in the Programme of Complementary Research in Fusion.
The views presented here do not necessarily represent those of the European
Commission.

\bigskip

\bigskip

\end{document}